\documentstyle[12pt,aasms4]{article}
\lefthead{Blackman \etal}
\righthead{Magnetic Fields in Planetary Nebulae}

\begin{document}

\title{Dynamos in Asymptotic-Giant-Branch Stars As the Origin 
of Magnetic Fields Shaping Planetary Nebulae}

\author{Eric G. Blackman\altaffilmark{1}, Adam Frank\altaffilmark{1}, 
J. Andrew Markiel\altaffilmark{2}, John H. Thomas\altaffilmark{1,3}, 
and Hugh~M.~Van Horn\altaffilmark{1}}

\altaffiltext{1}{Department of Physics and Astronomy and C. E. K. Mees
Observatory, University of Rochester, Rochester, NY 14627-0171, USA} 

\altaffiltext{2}{Department of Astronomy, University of Washington,
Seattle,
Washington 98195-1580, USA}

\altaffiltext{3}{Department of Mechanical Engineering, University of
Rochester, 
Rochester, NY 14627-0171, USA} 

%\keywords{ISM: bubbles -- MHD -- planetary nebulae -- stars: magnetic 
% fields -- stars: AGB and post-AGB}

\vfill
\eject

{\bf Planetary nebulae are thought to be formed when a slow wind 
from the progenitor giant star is overtaken by a subsequent fast wind as 
the star enters its white dwarf stage$^{1}$. 
The shock formed near 
the contact discontinuity between the two winds creates the relatively 
dense shell that forms the planetary nebula. A spherically symmetric wind 
produces a spherically symmetric nebula; however, over half 
of the known planetary nebulae are either bipolar or elliptical, rather 
than spherical $^{2}$.  A magnetic field may explain the launching 
and collimation of a bipolar outflow in a planetary nebula, but the origin 
of such a magnetic field has not been adequately explained. 
Here we show that a star on the asymptotic giant branch (AGB), which is the 
precursor of a planetary nebula core, can generate a strong magnetic field 
in a dynamo located at the core-envelope interface. The field is sufficiently 
strong to shape the bipolar outflow observed in planetary nebulae and may 
also explain the puzzlingly slow rotation of most white dwarfs via magnetic braking.}

One model for producing bipolar or elliptical planetary nebulae assumes 
that the slow wind from the progenitor star is denser near the equatorial plane 
than along the poles$^{3-6}$. Such an asymmetry can result if the progenitor 
is part of a close binary star system. However, the incidence 
of close binaries seems not to be sufficient to account for all asymmetric 
planetary nebulae.

Magnetic shaping is a promising mechanism for forming asymmetric planetary 
nebulae. A toroidal magnetic field can constrain the flow 
in the equatorial plane$^{7}$, redirecting it preferentially along the 
poles$^{8}$, while a dipole field can create a dense torus around the 
central star which then collimates the wind$^{9}$. The strength 
of a toroidal magnetic field in the wind will be increased in the shocked 
bubble, leading to collimation$^{10,11}$. Collimation might also occur 
close to the star through magneto-centrifugal processes$^{12}$. 

We suggest that the required magnetic field is generated by a dynamo in the 
central star during its AGB phase. The necessary ingredients for an alpha-omega
interface dynamo are very likely to be present near the outer boundary of the 
degenerate core of an AGB star. Contraction of the core and expansion of the 
envelope during this stage of evolution will tend to produce strong differential 
rotation regardless of the initial rotation profile on the main sequence. 
In the lower part of the deep, rotating convection zone the necessary alpha 
effect is provided by the helical convection itself or by magnetic instabilities. 
We note that a previous argument against significant dynamo action in an AGB 
star$^{13}$ was based on a model of a distributed dynamo operating solely within 
the slowly rotating envelope, a very different picture than the one we present 
here.  
	
To investigate the possibility of dynamo action in AGB stars, we first
estimate the rotation profile in a typical AGB star evolving from a rotating 
main-sequence star, and we then use this rotation profile in a self-consistent
nonlinear dynamo model to test for dynamo action. For a typical AGB star, we use
an evolutionary model of S. D. Kawaler (personal communication) for a star of mass 
$3 M_\odot$. We assume that this star is rotating uniformly on the main
sequence with angular velocity $1 \times 10^{-4} \: \rm{s}^{-1}$ (corresponding
to a surface rotation velocity $\sim \, 200 \: \rm{km \: s^{-1}}$, see ref. 14)
and that during the subsequent evolution the angular momentum of each spherical
mass shell is conserved. (As we expect some angular momentum exchange to occur in 
any real star, e.g. by magnetic braking, our model exaggerates the degree of 
differential rotation that may develop; nevertheless, we expect our model to be 
at least qualitatively correct.) The resulting rotation profile for this star at 
the tip of the AGB (Kawaler's model 1401, hereafter SDK 1401) has strong 
differential rotation, with the innermost core rotating faster 
and the outer envelope rotating much slower than the 
initial main-sequence rotation rate. The rotation profile in the neighborhood 
of the core-envelope interface is shown in Fig. 1. Our resulting rotation profile 
and thus our approach are strongly supported by observations that have been 
interpreted to imply core-envelope decoupling: observed fast rotation in 
horizontal-branch stars$^{15-17}$ and the existence of young, rapidly spinning 
pulsars$^{18}$.

The arrangement of differential rotation and convection in adjacent
layers in  this AGB star is similar to that in the Sun and strongly suggests
the possibility of an alpha-omega ``interface" dynamo. To test this
possibility we use a simple nonlinear alpha-omega dynamo model that we have used
previously to study dynamos in white dwarf stars$^{19}$. 
This model, based on a local analysis of the full mean-field dynamo
equations, assumes that the dynamo region consists of two adjacent, thin layers -- 
an inner layer in differential rotation and an outer layer of convection in which 
the alpha effect operates. The solar dynamo is thought to be of this type$^{20,21}$, 
and this configuration also corresponds 
roughly to the situation in our model AGB star. The nonlinear saturation of 
oscillatory dynamo modes in our model is caused by quenching of the alpha effect 
by the generated magnetic field. The free parameters in the model are calibrated by 
applying the model to the Sun and requiring it to produce a dynamo with the correct 
period (22 yrs) and appropriate amplitude (toroidal field strengths $\sim 10^4$ G). The 
calibrated model is then applied to the model AGB star. As input to our dynamo model, 
we obtain the following values from our model of 
differential rotation in SDK 1401: a differential rotation layer of thickness 
$4.6 \times 10^{10} \: \rm{cm}$ lying just below the convection zone 
(at radius $9.35 \times 10^{10} \: \rm{cm}$), with angular velocity dropping from 
$2 \times 10^{-5} \: \rm{s}^{-1}$ to $5 \times 10^{-6} \: \rm{s}^{-1}$ across this 
layer; and an alpha-effect layer of thickness $1 \times 10^{11} \: \rm{cm}$ lying 
just above the base of the convection zone, with typical convective velocity 
$1 \times 10^{5} \: \rm{cm \: s^{-1}}$ 
and density $6.6 \times 10^{-4} \: \rm{g \: cm^{-3}}$ (see Fig. 1).

For these input values for SDK 1401 we obtain an oscillatory dynamo 
with a period of $\sim \, 0.4$ yrs and a maximum mean toroidal field
strength of $\sim \, 5 \times 10^4$ G just below the convection zone.
(This result is robust, in the sense that it is not sensitive to 
moderate changes in the values of the input parameters.)
The ratio of thermal pressure to magnetic pressure just below the 
convection zone is $\beta \simeq  10^{3}$.
The field need not be volume-filling, and flux tube formation is very
likely as the magnetic field becomes subject to buoyancy instabilities. The
fraction of volume filled by flux tubes, and thus the surface-area covering 
fraction $f$ for long thin tubes, could be as low as $\beta^{-1}$ (ref. 22). 
The strength of a typical flux tube
in this case, determined by pressure balance across the tube, is  
$3\times 10^6$ G. The strength of such a tube upon rising to the top of the
convection zone ($R=5.5\times 10^{12}$ cm) would be $\sim 400$ G in pressure
balance. Surface magnetic fields of this strength
could be  associated with flares and coronal mass ejections from the AGB star
by analogy with the Sun$^{23}$.
These coronal mass ejections could be responsible for the non-axisymmetric, collimated
shapes with unpaired ejections of knots or bullets seen in some planetary 
nebulae$^{24,25}$.	

An average toroidal field strength of $5 \times 10^4$ G at the interface
represents $10^{41}$ ergs of magnetic energy in the volume of the shear
layer.  The rate of magnetic energy flow from the interface
layer due to buoyancy is $\sim (B^2/8\pi)(4\pi R_c^2) v_A$,
where  $v_A$ is the Alfv\'en speed in the layer and $R_c$ is the
radial location of the layer.
For $B= 5\times 10^4$ G, a density in the interface layer of $1.3
\times 10^{-3} \: \rm{g \: cm^{-3}}$,  and
$R_c = 9.35\times 10^{10}$ cm, this gives
an upward magnetic energy supply rate of $\sim  4 \times 10^{36}\:
{\rm erg \ s^{-1}}$. This number is comparable to the
turbulent energy dissipation rate in the convection zone, given by
$(\rho v^2)(v/l)({\mathcal V}_{conv})\sim 10^{36}\:
{\rm erg\ s^{-1}}$, when estimated for typical values, e.g.,  $\rho
\sim 10^{-4} \: \rm{g \: cm^{-3}}$
for the density, ${\mathcal V}_{conv}\sim 5 \times 10^{36}\: {\rm cm^3}$
for the volume, $v\sim 10^5 \: \rm{cm \: s^{-1}}$ for the turbulent
speed, and $l\sim 10^{11}$ cm for the characteristic eddy scale.
This means that the buoyancy-driven magnetic energy supply rate
is an upper limit on the supply rate to the corona, since a sizeable
fraction may be shredded and dissipated on its way up through the 
convection zone . The fraction that does make it to the corona would
be available for particle acceleration and coronal emission from the
AGB star. Some of this could  take the form of localized flares and
coronal mass ejections, as on the Sun.

Detection of X-rays from such a corona is unlikely while the star is 
in either the AGB stage or the proto-planetary-nebula stage due the 
high density of the wind. We estimate the optical depth for soft 
(1 KeV) and hard (10 KeV) X-rays to be 10 and $10^3$, respectively.
The X-rays could also produce a layer of highly ionized atoms near the 
star, but this would likely be enshrouded by much cooler gas in the 
stellar wind and thus be hard to detect. 
The deposition of energy into the corona may have consequences for the 
thermal and dynamical structure of the winds, which will need further 
investigation. We expect that the corona will be removed along with the 
bulk of the magnetic flux by the time the star reaches the mature 
planetary nebula phase.

The core of an AGB star can be expected to rotate very rapidly because of
contraction, as illustrated here by our model. However, the white
dwarf stars, which develop from AGB cores following the planetary 
nebula phase, generally rotate much more slowly than expected if angular
momentum conservation of the AGB core is assumed. 
This well-known problem$^{26}$ might
be resolved as a result of invoking the dynamo action proposed here;
the strong magnetic field generated by the AGB dynamo could produce
magnetic braking. This would occur during and after the time the AGB
star sheds its envelope.  Because of flux freezing, the field lines are 
drawn out with the envelope material but are anchored in the core. In this 
way the angular momentum of the core is transported outward to the 
envelope, and the core is spun down by a post-AGB MHD wind $^{12}$.  
During the braking period, bipolar and multipolar MHD winds could form
the 
shapes observed in some planetary nebulae as the stellar convective
envelope is shed.

The MHD wind luminosity then gives an upper limit to the kinetic
power$^{12}$. For an MHD wind from field lines anchored in the core, 
the luminosity can be as high as $L_{MHD}\sim 2\times 10^{38}(B/10^5 \:
{\rm G})^2(\Omega_c/2\times 10^{-5} \:  {\rm sec}^{-1})
(R_c/10^{11} \: {\rm cm})^3 \: {\rm erg \: s^{-1}}$, where $\Omega_c$
is the angular speed of the core material where the field lines are anchored,
and $R_c$ is the core radius. A $\sim 5$\% conversion of this is required 
to account for the energy loss rates of 
$10^{37}({\dot M}/6\times10^{21} \: {\rm g \: s^{-1}})(V/400 \: {\rm
km \: s^{-1}})^2$ $\rm{erg \: s^{-1}}$ characteristic of 
some proto-planetary nebulae$^{27}$, where 
$\dot M$ is the mass loss rate and $V$ is the wind
speed. The time scale for the MHD wind spin-down of the post AGB
star is then 
$\tau_s=140 (M_{c}/M_\odot)(\Omega_c/2\times 10^{-5} \: {\rm s^{-1}}) 
(B/10^4 \: {\rm G})^{-2}(R_c/10^{11} \: {\rm cm})^{-1}\ {\rm yr}$. 

The post-AGB wind, produced when the AGB star sheds its outer layers
and exposes the rapidly rotating, magnetized core, may be
strongly collimated by magneto-centrifugal processes$^{28}$.  
The degree to which a magnetized rotator will collimate
a wind driven off its surface can be expressed via a ``rotation''
 parameter$^{29}$ $Q$ given by
$Q/Q_\odot\simeq 4 (\psi_c/ 5\times 10^{26} \: {\rm G \: cm^2})(\Omega_c / 
2\times 10^{-5} \: {\rm s^{-1}})
({\dot M}/ 6\times 10^{21} \: {\rm g \: s^{-1}})^{-1/2}
(V / 400 \: {\rm km \: s^{-1}})^{-3/2}$,
where $\psi_c$ is the magnetic flux at large distances 
($\propto B R^2$), $V$ is the outflow speed, and $Q_\odot\sim 0.12$ is
the value for the solar wind (using
$V_\odot = 400 \: {\rm km \: s^{-1}}$, 
${\dot M}_\odot=1.6\times 10^{12} \: {\rm g \: s^{-1}}$,
$\Omega_\odot =3\times 10^{-6} \: {\rm s^{-1}}$, 
and $\psi_\odot=1.4\times 10^{22} \: {\rm G \: cm^{2}}$).  
We have scaled the flux to the upper limit
using $B\sim 5\times 10^4 \: {\rm G}$ and $R_c\sim 10^{11} \: {\rm cm}$
for our dynamo-produced field strength at the AGB core, and we 
have scaled the outflow parameters ${\dot M}$ and $V$ 
using proto-planetary nebulae values$^{27}$.
For $Q \ge 1$ the system is classified as a fast magnetic
rotator. The larger $Q$ is, the more strongly self-collimated
the outflow is$^{30}$.  We can see that for these parameters collimation 
greater than that in the solar wind is possible.  
If we instead use representative outflow parameters for later
stages of planetary nebulae$^{31}$, $\dot{M} =
5\times 10^{-7} M_\odot \: {\rm yr^{-1}}$ and $V = 1000 \: \rm{km \:
s^{-1}}$, 
we find $Q/Q_\odot\sim 15$, which implies that significant collimation
is possible.  Magnetic collimation may thus be intrinsic to these sources
and may not require shocks, as in some models$^{11}$. 

Our model leads us to predict that magnetic fields should be apparent 
in the winds of AGB stars and proto-planetary nebulae. Such magnetic 
fields should be detectable in some maser spots, and indeed have already 
been observed in at least one object$^{32}$ with field strengths 
consistent with our model (based on flux conservation), although this field
might be locally amplified by turbulence. We also predict that strongly 
collimated flows in proto-planetary nebulae should have signatures of 
ordered magnetic fields (e.g., polarization, Faraday rotation), reflecting 
the role of the magnetic field in their launching and collimating. At small 
distances from the central star, the field should be primarily polodial and 
parallel to the jet flow, while at large distances we would expect a 
dominant toroidal component (perpendicular to the outflow) as the hoop 
stresses take over the collimation.

Our dynamo model depends on rapid rotation of the AGB core. If a way can be 
found to measure the rotation rates of AGB cores, it could serve as an 
observational test of our model. Stars that rotate much more slowly than 
average on the main sequence would not be expected to produce significant 
dynamos at the AGB stage and hence would not be expected to produce 
collimated outflows by magnetic shaping. On the other hand, we expect those 
AGB stars that do have rapidly rotating cores to produce bipolar planetary 
nebulae.

As the rotation rate of the degenerate core is reduced by magnetic 
braking and the convective envelope is removed, the stellar dynamo will 
shut down. Some remnant field anchored in the core will survive even 
without a convection zone, although the convective envelope may not be 
removed completely. Indeed, white dwarfs do have thin surface
convection zones which can support a near-surface dynamo in the white 
dwarf itself$^{19}$.

In conclusion, we have demonstrated that dynamos are likely to
operate in AGB stars. As a star evolves off the AGB, the 
dynamo-generated magnetic field  will be strong enough to drive a
strong, self-collimating outflow and to slow the rotation of the core by
magnetic braking. Eruptions analogous to coronal mass ejections, expected
as a consequence of the dynamo activity, could produce asymmetric
structures in the wind. Thus our model opens up the possibility of 
constructing a new, self-consistent paradigm for planetary-nebula 
formation, beginning with AGB stars and ending with slowly rotating 
white dwarfs.

%\acknowledgments

We are grateful to Steve Kawaler and to Francesco D'Antona and Paolo 
Ventura for making available to us detailed tables of their evolutionary
models for AGB stars. This work was supported by the NSF, NASA, and DOE.

Correspondence and requests for materials should be addressed to J.H.T. 
(e-mail: thomas@astro.me.rochester.edu).

\vspace{0.5 in}
Figure 1. Internal rotation rate of our model $3.0\, M_\odot$ AGB star as a 
function of radius near the base of the convection zone. The angular 
velocity $\Omega$ is given both in units of s$^{-1}$ and in units of the 
initial (uniform) rotation rate $\Omega_0$
of the star when on the main sequence. The inner 
core has been spun up by contraction, while the outer layers rotate much 
more slowly than the initial rate due to the large expansion of the 
envelope. The dashed line indicates the location of the base of the 
convection zone; the inner dotted line indicates the inner boundary of 
the differential-rotation layer assumed in our dynamo model; and the outer 
dotted line indicates the outer boundary of the alpha-effect layer.


\begin{thebibliography}{}

\bibitem[Kwok, Purton, \& Fitzgerald 1978]{kwok78} 1. Kwok, S., Purton, 
    C. R. \& Fitzgerald, P. M. On the origin of planetary nebulae {\it
Astrophys. J.} {\bf 219}, L125-L127 (1978).

\bibitem [Manchado et al. 2000]{} 2. Manchado, A., Villaver, E.,
    Stanghellini, L. \& Guerrero, M.
    The morphological and structural classification of planetary nebulae.
    In {\it Asymmetrical Planetary Nebulae II. From Origins to
    Microstructures} (eds. J. H. Kastner, N. Soker \& S. Rappaport) 17-23 
    (ASP Conf. Ser. 199, 2000).

\bibitem[Balick 1987]{bali87} 3. Balick, B. The evolution of planetary
    nebulae. I - Structures, ionizations, and morphological sequences. 
    {\it Astron. J.} {\bf 94}, 671-678 (1987).

\bibitem[Soker \& Livio 1989]{soke89} 4. Soker, N. \& Livio, M. 
    Interacting winds and the shaping of planetary nebulae. 
    {\it Astrophys. J.} {\bf 339}, 268-278 (1989).

\bibitem[Mellema, Eulderink \& Icke 1991]{} 5. Mellema, G., Eulderink, F.
    \& Icke, V. Hydrodynamical models of aspherical planetary nebulae. 
    {\it Astron. Astrophys.} {\bf 252}, 718-732 (1991).

\bibitem[Icke, Balick, and Frank 1992]{icke92} 6. Icke, V., Balick, B. 
    \& Frank, A.  The hydrodynamics of aspherical planetary nebulae. II.
    Numerical modelling of the early evolution. 
    {\it Astron. Astrophys.} {\bf 253}, 224-243 (1992).

\bibitem[Gurzadyan 1969]{} 7. Gurzadyan, G. {\it Planetary Nebulae} (New York,
    Gordon and Breach, 1969).

\bibitem[Pascoli 1987]{pasc87} 8. Pascoli, G.  La nature des n\'ebuleuses
    plan\'etaires bipolaires. 
    {\it Astron. Astrophys.} {\bf 180}, 191-200 (1987).

\bibitem[Matt et al. 2000]{} 9. Matt, S., Winglee, D. \& Balick, B.
    Bipolar outflows without binarity. 
    {\it Astrophys. J.}, in press (2000).

\bibitem[Chevalier \& Luo 1994]{chev94} 10. Chevalier, R. A. \& Luo, D.
    Magnetic shaping of planetary nebulae and other stellar wind bubbles. 
    {\it Astrophys. J.} {\bf 421}, 225-235 (1994).

\bibitem[Garc\'ia-Segura et al.\ 1999]{garc00} 11. Garc\'ia-Segura, G., Langer,
    N., Rozyczka, M. \& Franco, J. Shaping bipolar and elliptical
    planetary nebulae: effects of stellar rotation, photoionization 
    heating, and magnetic fields. 
    {\it Astrophys. J.} {\bf 517}, 767-781 (1999).

\bibitem[Blackman et al 2000]{bfw00} 12. Blackman, E.G., Frank, A. \& Welch, C.
    Coupled MHD disk and stellar winds: Application to planetary nebulae.
    {\it Astrophys. J.}, submitted, astro-ph/0005288 (2000)

\bibitem[Soker \& Harpaz 1992]{sok92} 13. Soker, N. \& Harpaz, A. Can a single 
     AGB star form an axially symmetric planetary nebula? 
     {\it J. Astron. Soc. Pacific} {\bf 104}, 923-930 (1992). 

\bibitem[Cox 1999]{cox99} 14. Cox, A. N. {\it Allen's Astrophysical 
    Quantities}, 4th ed. (AIP Press, New York, 1999), 389.

\bibitem[Peterson 1983]{pet83} 15. Peterson, R. C. The rotation of 
	horizontal-branch stars. II. Members of the globular clusters M3, M5, 
	and M13. 
	{\it Astrophys. J.} {\bf 275}, 737-751 (1983).

\bibitem[Pinsonneault et al 1991]{pin91} 16. Pinsonneault, M., Deliyannis, 
    C. P. \& Demarque, P.
    Evolutionary models of halo stars with rotation. I. Evidence for 
    differential rotation with depth in stars.
    {\it Astrophys. J.} {\bf 367}, 239-252 (1991).

\bibitem[Behr et al 2000]{ber00} 17. Behr, B. B., Djorgovski, S. G., Cohen, 
    J. G., McCarthy, J. K., C\^ot\'e, P., Piotto, G. \& Zoccali, M.
    A new spin on the problem of horizontal-branch gaps: stellar 
    rotation along the blue horizontal branch of globular cluster M13.
    {\it Astrophys. J.} {\bf 528}, 849-853 (2000).

\bibitem[Livio \& Pringle 1998]{liv98} 18. Livio, M. \& Pringle, J. E.
    The rotation rates of white dwarfs and pulsars. 
    {\it Astrophys. J.} {\bf 505}, 339-343 (1998).

\bibitem[Thomas, Markiel, \& Van Horn 1995]{thom95} 19. Thomas, J. H., 
    Markiel, J. A. \& Van Horn, H. M.  Dynamo generation of magnetic
    fields in white dwarfs. 
    {\it Astrophys. J.} {\bf 453}, 403-410 (1995).

\bibitem[Parker 1993]{park93} 20. Parker, E. N.  A solar dynamo surface wave 
    at the interface between convection and nonuniform rotation. 
    {\it Astrophys. J.} {\bf 408}, 707-719 (1993).

\bibitem[Markiel \& Thomas 1999]{mark99} 21. Markiel, J. A. \&
    Thomas, J. H.  Solar interface dynamo models with a realistic
    rotation profile. 
    {\it Astrophys. J.} {\bf 523}, 827-837 (1999).

\bibitem[Blackman 1996]{b96} 22. Blackman, E.G. Overcoming the
    backreaction on turbulent motions in the presence of magnetic fields. 
    {\it Phys. Rev. Lett.} {\bf 77}, 2694-2697 (1996).

\bibitem[Pick 1999]{Pic00} 23. Pick M., Coronal mass ejections, in 
    {\it Ninth European Meeting on Solar Physics: Magnetic Fields and 
    Solar Processes} (ed. A. Wilson) 183 (ESA SP Series SP-448, 1999). 

\bibitem[Trammell 2000]{tr00} 24. Trammell, S. R. Hubble Space Telescope
    imaging of the young planetary nebula AFGL 618. In 
    {\it Asymmetrical Planetary Nebulae II. From Origins to
    Microstructures} (eds. J. H. Kastner, N. Soker \& S. Rappaport) 147-150 
    (ASP Conf. Ser. 199, 2000).

\bibitem[Sahai 2000]{Sa00} 25. Sahai, J., Imaging of PPNe with HST. In 
    {\it Asymmetrical Planetary Nebulae II. From Origins to
    Microstructures} (eds. J. H. Kastner, N. Soker \& S. Rappaport) 209-216 
    (ASP Conf. Ser. 199, 2000).
           
\bibitem[Koester et al 1998]{Koe98} 26. Koester, D., Dreizler, S., Weidemann, V. 
     \& Allard, N. F. Search for rotation in white dwarfs. 
     {\it Astron. Astrophys.} {\bf 338}, 612-622 (1998).

\bibitem[Alcolea et al 2000]{Alc00} 27. Alcolea, J., Bujarrabal, V., 
    Castro-Carrizo, A, Sanchez Contreras, C., Neri, R. \& Zweigle, J. 
    Molecular line observations of proto-planetary nebulae.
    In {\it Asymmetrical Planetary Nebulae II. From Origins to
    Microstructures} (eds. J. H. Kastner, N. Soker \& S. Rappaport) 347-354 
    (ASP Conf. Ser. 199, 2000).

\bibitem[K\"onigl \& Pudritz 2000]{PK00} 28. K\"onigl, A. \& Pudritz, R. 
    Disk winds and the accretion outflow connection. 
    In {\it Protostars and Planets IV} (eds. Mannings, V.,  Boss, V, A. \& 
    Russel, S.) 759-791 (U Arizona Press, Tucson, 2000).

\bibitem[Tsinganos \& Bogovalov 2000]{ts00} 29. Tsinganos, K. \& Bogovalov
    S. Magnetic collimation of solar and stellar winds. 
    {\it Astron. Astrophys.}, in press (2000).

\bibitem[Lery et al 1999]{le99} 30. Lery, T., Heyvaerts, J., Appl, S. \&
    Norman, C. Outflows from magnetic rotators. 
    {\it Astron. Astrophys.} {\bf 337}, 603-624 (1998).

\bibitem[Balick et al. 1998]{ba98} 31. Balick, B, Alexander, J., Hajian, A.,
    Terzian, Y, Perinotto, M. \& Patriarchi, P. FLIERs and other 
    microstructures in planetary nebulae. IV. Images of elliptical PNs 
    from the Hubble Space Telescope. 
    {\it Astron. J.} {\bf 116}, 360-371 (1999).

\bibitem[Palen \& Fix 2000]{pal00} 32. Palen, S. \& Fix, J. D. Models of OH maser 
     variations in U Herculis. {\it Astrophys. J.} {\bf 531}, 391-400 (2000).


\end{thebibliography}
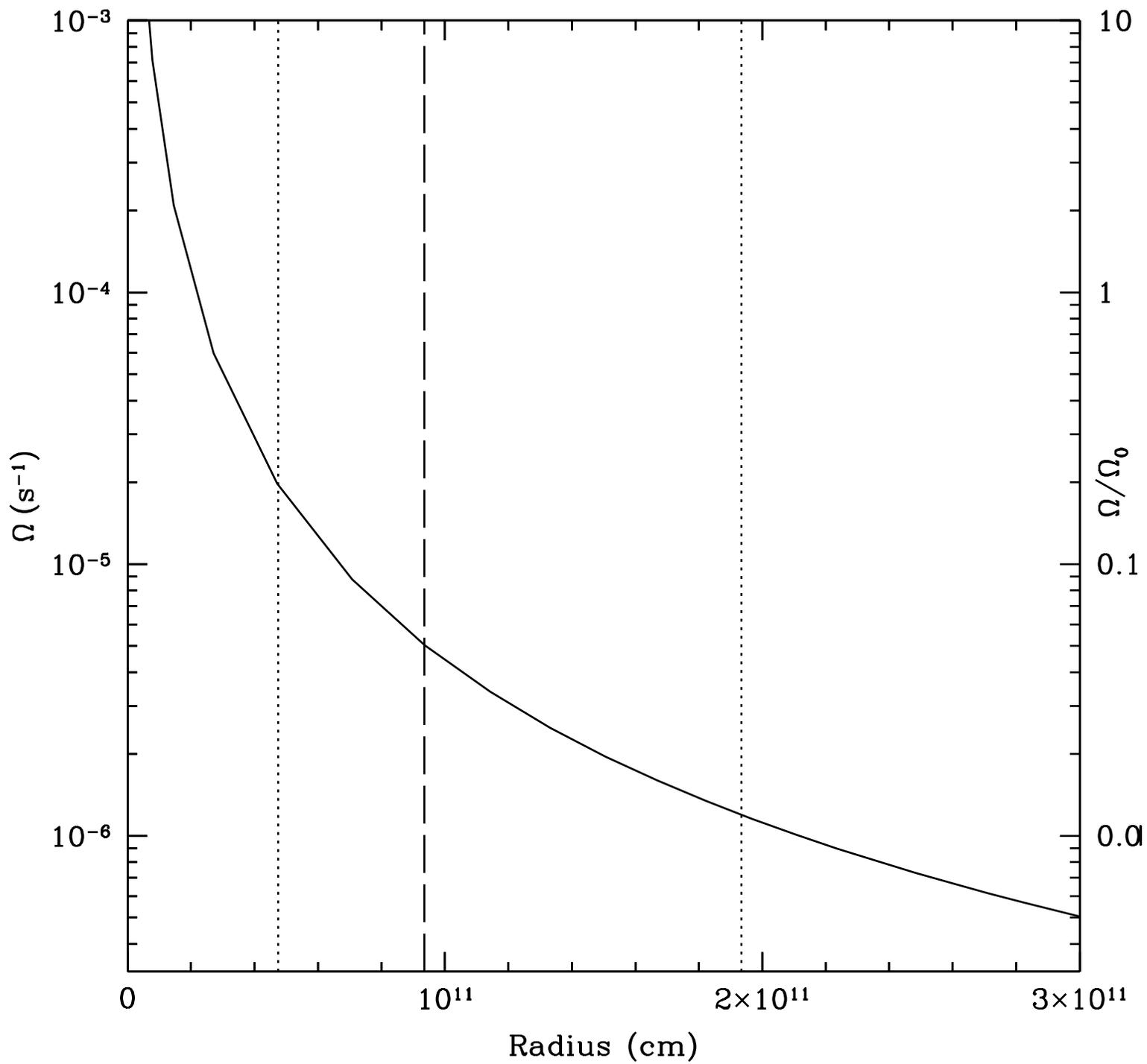
\end{document}